      [1999/12/01 v1.4c Il Nuovo Cimento]
\documentclass{cimento}

\usepackage{graphicx} 
\title{The GRB 030328 host: another case of a blue starburst galaxy\thanks{Based on data taken at the 2.2-m and 3.5-m telescopes of
the Centro  Astron\'omico Hispano Alem\'an  de Calar Alto, operated  by the
Max Planck  institute of Heidelberg and Centro  Superior de Investigaciones
Cient\'{\i}ficas.}}
\author{J.~Gorosabel\from{1},
M. Jel\'{\i}nek\from{1},
A.~de Ugarte Postigo\from{1},
S.~Guziy\from{1},
\atque
A.J.~Castro-Tirado\from{1}}

\instlist{\inst{1} Instituto de Astrof\'{\i}sica de Andaluc\'{\i}a (IAA-CSIC),
           P.O. Box 03004, E-18080 Granada, Spain}

\PACSes{\PACSit{95.75.De}{Photography and photometry}
        \PACSit{98.70.Rz}{gamma-ray sources; gamma-ray bursts}}

\begin{document}

\maketitle

\begin{abstract}
  We present for  the first time the detection  of the GRB~030328 host
  galaxy in four optical bands  equivalent to $UBRI$.  The host galaxy
  spectral  energy distribution  is consistent  with a  low extinction
  ($E(B-V)  <   0.21$)  starburst  galaxy.    The  restframe  $B$-band
  magnitude of the host is $M_{B}\sim-20.4$.

\end{abstract}

\section{Introduction}

A long gamma-ray  burst (GRB) was detected on 2003  March 28 at 11:20:58.34
UT by  the HETE-2  spacecraft \cite{Vill03}. The  gamma-ray event  showed a
duration of approximately 100 seconds in  the 30-400 keV energy band, so it
belongs to  the category of ``long-duration'' GRBs  \cite{Kouv93}.  The GRB
peak flux over 5.2 seconds was $7.3 \times 10^{-7}$ erg cm$^{-2}$ s$^{-1}$,
and the  fluence was  $3.0 \times 10^{-5}$  erg cm$^{-2}$ in  the mentioned
energy band.  Optical  images taken $\sim1.3$ hours after  the GRB revealed
the  associated  optical   afterglow  (OA)  \cite{Pete03}.   Subsequent  low
resolution spectroscopy determined a redshift value of $z=1.52$ for the OA
\cite{Mart03,Rol03}.   Nevertheless  no  host  galaxy  detection  has  been
reported to date.

Throughout,   we   assume   a   cosmology   where   $\Omega_{\Lambda}=0.7$,
$\Omega_{M}=0.3$  and   $H_0=72$  km  s$^{-1}$   Mpc$^{-1}$.   Under  these
assumptions, the  luminosity distance of  GRB~030328 is $d_l=10.8$  Gpc and
the look-back time is 9.05~Gyr ($\sim69.2$ \% of the present Universe age).

\section{Observations}

In  the present  paper we  report optical  multicolour imaging  of  the GRB
030328 host galaxy.  The observations  were carried out with the Calar Alto
(CAHA) 2.2-m telescope $\sim$1 year after the GRB, when the contribution of
the  OA was  negligible.  The  data were  collected with  the  BUSCA camera
\cite{Reif00},  which allows  simultaneous imaging  in four  optical bands.
The  four channels  (named  $C1$, $C2$,  $C3$  and $C4$)  are not  standard
filters,  so  they were  calibrated  by  observing the  spectro-photometric
standard star  GD153 \cite{Bohl95} at the  same airmass to that  of the GRB
field   (information  on  the   four  BUSCA   channels  are   described  in
\cite{Goro05}).  $C1$ is similar the Johnson $U$-band, $C2$ is a transition
between  Johnson's  $B$  and  $V$  filters.   $C3$ is  very  close  to  the
$R_c$-band and $C4$ is identical to the $I_c$ filter.  The covered field of
view  (FOV) was  $12.0^{\prime}  \times 12.0^{\prime}$,  and the  resulting
pixel scale $0.35^{\prime\prime}$/pix.

\begin{figure}
\begin{center}
\resizebox{7.8cm}{!}{\includegraphics{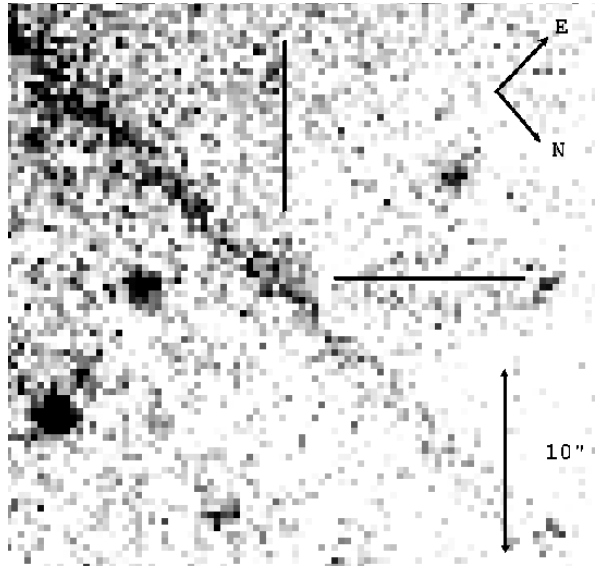}}
\caption{\label{Fig1} The tick-marks indicate the GRB 030328 host galaxy. 
The image have been created co-adding the images of the four channels.
As can  been seen, a spike  originated at a bright  $R\sim$8 mag field
star (out  of the FOV toward  the left upper corner)  goes through the
host galaxy.}
\end{center}
\end{figure}

The photometric  and astrometric analysis  of the field was  complicated by
the presence of two  bright field stars ($R\sim8$) located $\sim3^{\prime}$
south from the  OA position. In order to mitigate  the high background they
generated,  the BUSCA  camera  was  rotated 45  degrees  and the  telescope
pointing was  shifted that both stars were  out of the FOV.  Even with this
configuration  some degree of  contamination was  unavoidable, and  a spike
coming from the star crossed the OA position (see Fig.~\ref{Fig1}).

Reduction was based on standard procedures running under IRAF\footnote{IRAF
is distributed  by the National Optical Astronomy  Observatories, which are
operated  by the  Association of  Universities for  Research  in Astronomy,
Inc., under  cooperative agreement  with the National  Science Foundation.}
and the photometry  on PHOT.  To account for  the spike contamination three
apertures were  used; two positioned along  on the spike  bracketing the OA
equidistantly, and the third one on the OA. The fluxes of the two apertures
were  averaged  and  subtracted from  the  central  one  at the  OA.   This
procedure only integrated the OA flux.  The process was repeated moving the
two   apertures  along   the  spike,   giving  the   magnitude   errors  of
Table~\ref{table1}.

\begin{table}
\begin{center}
\caption{Optical observations carried out for the GRB~030328 host galaxy.
The host galaxy AB Magnitudes have been corrected for Galactic reddening
\cite{Schl98}.}
\begin{tabular}{lcccc}
\hline
\hline
Date (UT)  &  Filter & T$_{\rm exp}$&  Seeing        & Magnitude\\
March 2004 &         &    (s)       & (arcsec)       &   (AB)   \\
\hline
\hline
23.9549--24.1641& $C1$&24$\times$600& 1.7$^{\dagger}$&24.98$\pm$0.30\\
25.0366--25.1513&     &14$\times$600&                &     \\
\hline
23.9549--24.1641& $C2$&24$\times$600& 1.5$^{\dagger}$&24.72$\pm$0.35\\
25.0366--25.1513&     &14$\times$600&                &\\
\hline
23.9549--24.1641& $C3$&24$\times$600& 1.4$^{\dagger}$&24.39$\pm$0.35\\
25.0366--25.1513&     &14$\times$600&                &\\
\hline
23.9549--24.1641& $C4$&24$\times$600& 1.2$^{\dagger}$&24.42$\pm$0.40\\
25.0366--25.1513&     &14$\times$600&                &\\
\hline
\hline
\multicolumn{5}{l}{$\dagger$ Full width have maximum (FWHM) of the co-added image.}\\
\hline
\label{table1}
\end{tabular}
\end{center}
\end{table}

\section{Results}

The astrometric solution for each of  the four co-added images was based on
$\sim60$ USNO-A2.0  stars.  The final host  coordinates, obtained averaging
the four  channels, are; $\alpha_{J2000}=12^{h}10^{m}48.37^{\prime\prime}$,
$\delta_{J2000}=-09^{\circ}20^{\prime}51.0^{\prime\prime}$.              The
astrometric  error  for  each  coordinate  is  $0.7^{\prime  \prime}$,  not
accounting   for  the   systematic   error  of   the  USNO-A2.0   catalogue
($\sim0.25^{\prime\prime}$,  see \cite{Assa01}). The  main source  of error
comes from the uncertainty on the centroid position due to the bright spike
contamination. The  inferred position is  consistent with the  one reported
for the OA \cite{Pete03}.

\begin{figure}
\begin{center}
\resizebox{8.5cm}{!}{\includegraphics{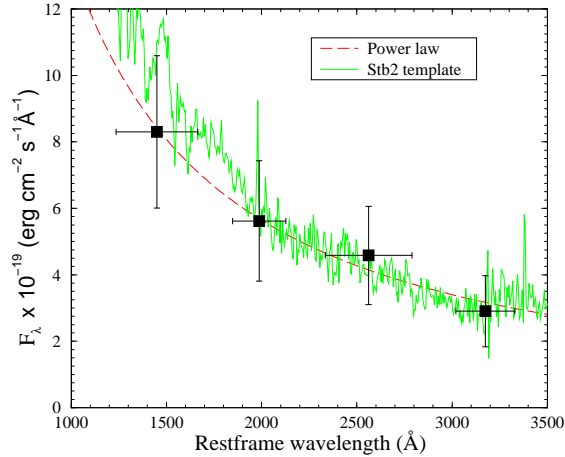}}
\caption{\label{Fig2} The restframe SED of the GRB 030328 host galaxy once
the   flux   densities  have   been   corrected   for  Galactic   reddening
\cite{Schl98}. The  squares show  the host flux  densities corresponding
to the four BUSCA channels. The horizontal error bars represent the FWHM of
the BUSCA filters  (measured in the host restframe).   The long dashed line
shows the  satisfactory power  law fit obtained  with $\beta=  -1.25$.  The
spiky  line  displays  the  best  fit  obtained  with  the  K96  templates,
corresponding to a  starburst galaxy with $0.11 <  E(B-V) < 0.21$ (template
Stb2).}
\end{center}
\end{figure}

Fig.~\ref{Fig2}  shows the  SED  of the  GRB  030328 host  galaxy once  the
observed  flux  densities  have   been  corrected  for  Galactic  reddening
($E(B-V)=0.047$, \cite{Schl98}). As can be  seen a power law fit provides a
satisfactory fit  to the  SED ($\chi^2/d.o.f =  0.08$) yielding  a spectral
index value of $\beta=-1.25\pm0.54$ ($F_{\lambda} \sim \lambda^{\beta}$).

In order  to infer information  on the host  extinction the SED  was fitted
with  the  empirical  galaxy  templates  described  in  \cite{Kinn96}  (K96
hereafter).  The  observed K96  templates can be  grouped into  seven sets:
Bulge (B), E,  S0, Sa, Sb, Sc and Stb galaxies.   Additionally, the Stb K96
templates are subdivided into six  classes depending on their extinction by
increasing $E(B-V)$ in steps of 0.1 (from Stb1 to Stb6; see more details in
K96). The extinction for  the B, E, S0, Sa, Sb, and  Sc templates is a free
parameter.

The   best  fit   is   obtained  with   the   blue  Stb1   and  Stb2   SEDs
($\chi^2/d.o.f\sim0.1$), which  are defined to have  the lowest extinctions
among  the K96  starburst templates  ($0 <  E(B-V) <  0.21$).  The  rest of
redder  templates  yield  worse   results.   Thus,  we  conclude  that  the
GRB~030328 host  is likely  a blue low  extinction starburst  galaxy.  This
result  agrees with  the conclusion  by \cite{Chri04},  who found  that GRB
hosts  are  similar to  the  blue starburst  galaxies  of  the Hubble  Deep
field. For  the Stb1 and Stb2  SED fits the absolute  $B$-band magnitude of
the host  is $M_{B}  \sim -20.4$.   No information can  be obtained  on the
stellar population  age since  the 4000~\AA~ jump  is beyond  our restframe
wavelength range.

\acknowledgments
This research is partially supported by the Spanish Ministry of Science and
Education   through  programmes   ESP2002-04124-C03-01   and  AYA2004-01515
(including FEDER funds).  JG acknowledges  the support of a Ram\'on y Cajal
Fellowship from the Spanish Ministry of Education and Science.


\begin{thebibliography}{0}
\bibitem{Vill03} \BY{Villasenor~J., Crew~G., Vanderspek~R., et al.} 
                 \IN{GCN Circ.}{1978}{2003}{}
\bibitem{Kouv93} \BY{Kouveliotou C., Meegan C.A., Fishman G.J., et al.} 
                 \IN{ApJL}{413}{1993}{101}
\bibitem{Pete03} \BY{Peterson~B.A. \atque Price~P.A.} 
                 \IN{GCN Circ.}{1974}{2003}{}
\bibitem{Mart03} \BY{Martini~P., Garnavich~P. \atque Stanek~K.Z.}
                 \IN{GCN Circ.}{1979}{2003}{}
\bibitem{Rol03}  \BY{Rol~E., Vreeswijk~P. \atque Jaunsen~A.}
                 \IN{GCN Circ.}{1981}{2003}{}
\bibitem{Assa01} \BY{Assafin~M., Andrei~A.H.,R.~Vieira Martins, et al.}
                 \IN{ApJ}{552}{2001}{380}
\bibitem{Reif00} \BY{Reif~K., Poschmann~H., Bagschik~H., et al.}
  in \TITLE{Optical  Detectors for Astronomy: State-of-the-Art  at the Turn
                  of  the Millenium. 4th  ESO CCD  Workshop, 1999,  held in
                  Garching,  Germany},  edited  by  \NAME{Amico~P.  \atque
                  Beletic~J.W.}  (Kluwer Academic Publisher) 2000, pp. 143.

\bibitem{Bohl95} \BY{Bohlin~R.C., Colina~L. \atque Finley~D.S.}
                 \IN{AJ}{110}{1995}{1316}                        

\bibitem{Goro05} \BY{Gorosabel~J., P\'erez-Ram\'{\i}rez~M.D., Sollerman~J., et al.}
                 \IN{A\&A}{submitted}{2005}{}

\bibitem{Schl98}  \BY{Schlegel~D.J., Finkbeiner~D.P. \atque David~M.}  
		  \IN{ApJ}{500}{1998}{525}                        

\bibitem{Kinn96}  \BY{Kinney~A.L., Calzetti~D., Bohlin~R.C., et al.}  
		  \IN{ApJ}{467}{1996}{38}                        

\bibitem{Chri04} \BY{Christensen~L., Hjorth~J. \atque Gorosabel~J.}
		  \IN{ApJ}{425}{2004}{913}                        

\end{thebibliography}
\end{document}